\documentclass[12pt]{iopart}
\usepackage{iopams}
\usepackage{setstack}
\usepackage[latin1]{inputenc}
\usepackage[english]{babel}
\usepackage{cite}
\usepackage[pdftex]{graphicx}
\usepackage[pdftex]{hyperref}
\usepackage{pdflscape}

\begin{document}

\title[Rb-Cs cold collision]{Determination of the scattering length for Rb-Cs X$^{1}\Sigma ^{+}$ ground electronic state using a variational method}

\author{M. N. Guimar\~aes\footnote{\mailto{mng@ufba.br}}\, and F. V. Prudente\footnote{\mailto{prudente@ufba.br}}}
 
\address{Instituto de F\'{\i}sica, Universidade Federal da Bahia, 40170-115, 
Salvador, Bahia, Brazil.} 

\begin{abstract}

We performed the calculation of the scattering length for the elastic collision between the rubidium and cesium atoms. For this we applied a variational procedure based on the $R$-matrix theory for unbound states employing the finite element method (FEM) for expansion of the wave-function in terms of a finite set of local basis functions. The FEM presents as advantages the possibility of the development of a efficient matrix inversion algorithm which significantly reduces the computation time to calculate the $R$ matrix. We also tested a potential energy curve  with spectroscopic accuracy obtained before from a direct adjustment procedure of experimental data of the $X^{1}\Sigma^{+}$ state based on genetic algorithm. The quality of our result was evaluated by comparing them with several ones previously published at literature.

\end{abstract}

\pacs{30.50.Cx; 02.70.Dh; 03.65.Nk; 31.50.Bc}

\submitto{arXiv}

\section{Introduction}

Especially in recent years, researches on ultracold atoms have led to important discoveries in atomic physics, notably the observation of Bose-Einstein condensation in gases of alkali atoms \cite{weiner(2003)}. The collision with a second atom has been attracting interest because, among other things, opens the possibility of systematically cooling one of the atomic species. From the theoretical viewpoint, ultracold collisions involve very large interatomic distances, typically in the order of thousands of bohr, which makes difficult the numerical solution of the scattering problem. In the study of elastic collision between two atoms we consider the knowledge of the potential energy curve for short and long internuclear distances, and the asymptotic physical quantities are expressed in terms of $S$-matrix and phase shift obtained using the partial waves decomposition. In particular, the scattering length, in sign and magnitude, enables to know the character of the interaction between atoms at very low energies.

Among the motivations for studying the cold and ultracold collision between the rubidium and cesium atoms is that the RbCs molecule has been subject of several recent spectroscopy studies \cite{shimasaki-et-al(2015),molony-et-al(2014),takekoshi-et-al(2014),bruzewicz-et-al(2014),zuters-et-al(2013),takekoshi-et-al(2012),yang-et-al(2012),docenko-et-al(2011)}. In fact, it is the heaviest heteronuclear alkali diatomic molecule with the largest permanent dipole moment being a candidate for experiments with dense ultracold ensembles. The Bose-Einstein condensate has been successfully obtained for both atomic species, and cold, and ultracold, molecules, specially for the X$^{1}\Sigma ^{+}$ ground eletronic state, have been produced, as well as, collisional properties have been calculated. Studies of ultracold atomic mixtures could also be important in applications in quantum computing.

In the theoretical studies of cold and ultracold collisions the variational methods have been unusual because of large matrices produced due to the very long range of potential. However, these procedures have proven to be a very powerful tool in developing numerical solutions for problems involving quantum scattering processes \cite{adhikari(1998)}. There are a broad range of approaches leading to many different problems. In all of them the solution is expanded in terms of known basis functions and the coefficients of expansion are determined by solving a set of linear algebraic equations, yielding at last the scattering observables. In particular, a procedure that has contributed to the recent progress is the $R$-matrix method originally proposed in 1947 by Wigner and Eisenbud \cite{wigner-eisenburg(1947)} in the nuclear physics context, but which has been applied to several problems in atomic and molecular physics \cite{burke(2011)}. The variational principle via the $R$-matrix theory is formulated so as to lead to a problem of matrix inversion yielding as a result the $R$ matrix, which is then connected to the $S$ matrix.

The success of the variational calculation will depend on the correct choice of basis set; if it is appropriate to the problem, then the results will be accurate and it will be required a lower computational effort for execution of problem solution. For this purpose, a very accurate procedure is the finite element method (FEM). The FEM has been widely used in the analysis of engineering problems, but over the years it has been applied in both study of bound states as for scattering processes of quantum systems \cite{ram-moham(2002),pask-et-al(2001),soares-prudente(1994),prudente-soares2(1999),linderberg(1989),jaquet-schnupf(1992)}. As a variational approach, in addition to provides a means to systematically improve the accuracy in the calculations in a natural way, the FEM has as advantages the possibility of the development of a efficient matrix inversion algorithm which significantly reduces the computation time to calculate the $R$ matrix.

In this paper we performed the calculation of the scattering length for the elastic collision between the Rb and Cs pairs using the variational $R$-matrix and finite element methods which are presented in section \ref{methodology}. We also tested a potential energy curve with spectroscopic accuracy obtained before from a direct adjustment procedure of experimental data of the $X^{1}\Sigma^{+}$ state based on genetic algorithm. The results are presents in section \ref{results}.

\section{Methodology}\label{methodology}

\subsection{Variational $R$-matrix method}

According to the variational formalism, the problem of solving the radial Schrödinger equation for a system composed by two atoms is equivalent to finding the solutions of the following functional of energy
\begin{equation}\label{functional}
J_{l}[\chi _{l}]=\int dr\chi _{l}^{\ast }(r)\left\lbrace -\frac{\hbar ^{2}}{2\mu }\frac{d^{2}}{dr^{2}}+V_{l}^{ef}(r)-E\right\rbrace \chi _{l}(r),
\end{equation}
where $l$ designates quantum number associated with the angular momentum, $\mu $ is the reduced mass, $\chi _{l}(r)$ is the radial solution and $V_{l}^{ef}(r)=V(r)+\frac{\hbar ^{2}l(l+1)}{2\mu r^{2}}$ is the effective potential, where $V(r)$ the potential energy curve.

The $R$-matrix method relates the wave function with its normal derivative on the boundary surface between the asymptotic and interaction regions defined by the point $r=r_{max}$. In particular, this method in variational form specifies that
\begin{equation}\label{r-matrix condition}
\chi _{l}(r_{max})=R_{l}\left. \frac{d\chi _{l}(r)}{dr}\right\vert _{r_{max}},
\end{equation}
where $R_{l}$ is the $R$ matrix for given $l$.

In order to have $R_{l}$, we expand the radial wave function in a finite set of basis functions,
\begin{equation}\label{expansion}
\chi _{l}(r)=\underset{j=1}{\overset{p}{\sum }}c_{j}^{l}f_{j}(r),
\end{equation}
replace this expansion, and the condition \eref{r-matrix condition}, in the functional  \eref{functional} and impose the stationarity condition obtaining the $R$ matrix for a collision process with one asymptotic channel:
\begin{equation}\label{r-matrix}
R_{l}=\frac{\hbar ^{2}}{2\mu }\underset{i=1}{\overset{p}{\sum }}\underset{j=1}{\overset{p}{\sum }}f_{i}^{\ast }(r_{max})\left\{ \mathbf{B}_{l}^{-1}\right\} _{ij}f_{j}^{\ast }(r_{max}),
\end{equation}
with $\mathbf{B}_{l}\equiv\mathbf{H}_{l}-E\mathbf{O}$ where $\mathbf{H}_{l}$ and $\mathbf{O}$ are the Hamiltonian and Overlap matrices, respectively.

The relation between $R_{l}$ and the scattering matrix for the $l$-th partial wave, $S_{l}$, can be achieved using the continuity of the function at point $r=r_{max}$ and written in the form
\begin{equation}
S_{l}=e^{2i\delta _{l}}=\frac{\left[ 1+ikR_{l}\right] }{\left[ 1-ikR_{l}\right] }e^{-2i\left( kr_{max}-\frac{l\pi }{2}\right) },
\label{s-matrix}
\end{equation}
where $\delta _{l}$ is the phase shift and $k=\sqrt{2\mu E/\hbar ^{2}}$. The scattering length, $a$, is related to the cross section for very small energies ($k^{2}\rightarrow 0)$ wherein only waves with $l=0$ contribute to scattering and it is given by
\begin{equation}
-\frac{1}{a}=\underset{k\rightarrow 0}{\lim }\left[ k\cot \delta _{0} \right].
\label{scattering length}
\end{equation}
On the other hand, for small value of wave number, $k$, the phase shift, for $l=0$, can be connected with the scattering length by the effective range expansion:
\begin{equation}
k\cot \delta _{0}=-\frac{1}{a}+\frac{1}{2}r_{c}k^{2}+O(k^{4}),
\label{effective range expansion}
\end{equation}
where $r_{c}$ is the effective range \cite{bethe(1949),blatt(1948),blatt-jackson(1949)}. Therefore, the scattering length, as well as the effective range, can also be calculated by equation \eref{effective range expansion}.

A feature of the $R$ variational formalism for scattering process is that, despite the $S$ matrix be complex, the $R$ matrix is real and ensures its symmetry and unitarity. On the other hand, a large computational effort is required the inversion of the matrix $\mathbf{H}_{l}-E\mathbf{O}$ in order to obtaining $R_{l}$ and such effort increases with the cube of the matrix order. But, this can be overcome by the finite element method which has great advantages when used to expand the function \eref{expansion} and solve the equation \eref{r-matrix}.

\subsection{Finite element method}

The finite element method (FEM) applied to the current problem consists basically in divide the integration interval $[0,r_{max}]$ into $N_{e}$ elements, being the $i$th element defined in the range of $r_{i-1}$ to $r_{i}$ with $r_{0}=0$ and $r_{N_{e}}=r_{max}$, and expand the radial wave function as follows
\begin{equation}\label{fem expansion}
\chi (r)=\overset{N_{e}}{\underset{i=1}{\sum }}\overset{n_{i}}{\underset{j=0}{\sum }}c_{j}^{i}f_{j}^{i}(r),
\end{equation}
where the parameter $n_{i}$ is the highest order of the basis functions associated with the $i$th element, $f_{j}^{i}(r)$ is the $j$th basis function of the same element and {$c^{i}_{j}$} are the expansion
coefficients. Here, the $l$-index associated with the angular momentum has been omitted.

The functions \{$f_{j}^{i}(r)$\} satisfy the following property
\begin{equation}\label{FEM property}
f_{j}^{i}(r)=0\quad \mbox{if} \quad r\notin \lbrack r_{i-1},r_{i}].
\end{equation}
In particular, we utilizes, into each element, two interpolant functions, $I^{i}_{1}\equiv f_{0}^{i}(r)$ and $I^{i}_{2} \equiv f_{n_{i}}^{i}(r)$, and polynomial shape functions, $\mathcal{P}^{i}_{j}\equiv f_{j}^{i}(r)$ with $j=1,\cdots ,n_{i}-1$ (see Ref.~\cite{guimaraes-prudente(2005)} for details).
These basis functions have an important feature of that only the basis function $I^{N_{e}}_{2} \equiv f_{n_{N_{e}}}^{N_{e}}(r)$ is non-null on the last node of the mesh:
\begin{equation}\label{elements relationship}
f_{j}^{i}(r_{N_{e}})=\left\{ 
\begin{array}{cc}
1 & \mbox{for } i=N_{e} \mbox{, } j=n_{N_{e}} \\ 
0 & \mbox{otherwise }
\end{array}
\right. .
\end{equation}

Because of equation \eref{FEM property}, the elements of matricial representation of $\widehat{B}$ operator have the following property:
\begin{equation}\label{matrix representation property}
\{\mathbf{B}\}_{jj^{\prime }}^{ii^{\prime }}=\underset{a}{\overset{b}{\int }}{\rm d} rf_{j}^{i}(r)\widehat{B}\ f_{j^{\prime }}^{i^{\prime }}(r)=0 \quad , \quad \forall i\neq i^{\prime}.
\end{equation}
This leads to matrices with an interesting block tridiagonal structure.
Due to properties \eref{matrix representation property} and \eref{elements relationship}, when we utilize the FEM for expanding the radial wave functions \eref{expansion} the $R$ matrix \eref{r-matrix} is written in the following form
\begin{equation}
R_{l}=\frac{\hbar ^{2}}{2\mu }\left\lbrace \mathbf{B}_{l}^{-1}\right\rbrace ^{N_{e}+1 \, N_{e}+1},
\label{FEM r-matrix}
\end{equation}
where $\mathbf{B}_{l}\equiv \mathbf{H}_{l}-E\mathbf{O}$ and the superscript indices in $\mathbf{B}_{l}^{-1}$ represent its last block.

Therefore, we need to know only the last block of the inverse of $\mathbf{H}_{l}-E\mathbf{O}$ matrix to obtain $R_{l}$.
This can be done efficiently utilizing an algorithm developed by Prudente and Soares Neto \cite{prudente-soares(1999)} aiming to calculate only the last block of inverse matrix. It reduces significantly both the computational time to invert the matrix as the memory required to store it on the computer and is demonstrated in details in reference \cite{guimaraes-prudente(2011)}.

\section{Results}\label{results}

In this section, data are presented to the elastic scattering of the cesium (${}^{133}$Cs) by rubidium (${}^{85}$Rb and ${}^{87}$Rb) atoms at temperatures close to absolute zero, interacting via the ground state ($X^{1}\Sigma ^{+}$) of the RbCs molecule. Specifically, we have presented scattering length and effective range calculations for the potential energy curve (PEC) obtained by Almeida \textit{et al} \cite{almeida-et-al(2011)} with spectroscopic accuracy from a direct adjustment procedure of experimental data based on the genetic algorithm. All calculations were performed using a computational implementation in Fortran based on variational $R$-matrix method and the finite element method (FEM). Shortly, for a partial wave with $l=0$, we solve the matrix inversion problem, given by the equation \eref{r-matrix} or \eref{FEM r-matrix},  using the matrix inversion technique \cite{guimaraes-prudente(2011)}. Having computed the $R$ matrix, we then calculate the phase shift, $\delta _{0}$, from the equation \eref{s-matrix}. The scattering length, $a$, is obtained by its definition given by equation \eref{scattering length}. The FEM uses the same polynomials order for all mesh elements (\textit{i.e.}, $n_{i}=const.,\forall i$) and the dimension of the $\mathbf{B}_{l}=\mathbf{H}_{l}-E\mathbf{O}$ matrix is $(N_{e}\cdot n_{i}+1)\times (N_{e}\cdot n_{i}+1)$ with its last block having dimension $1$.

The analytic function used to represent the $X^{1}\Sigma ^{+}$ electronic state PEC of RbCs molecule is as follows
\begin{equation}
V(r)=\left( \sum _{i=1}^{5}a_{i}r^{i-2}\right)e^{-(a_{6}r+a_{7}r^{2})}-\sum _{k=3}^{5}f_{2k}(a_{8}r)\frac{C_{2k}}{r^{2k}},
\label{PEC Almeida}
\end{equation}
where
\begin{equation*}
f_{2k}(a_{8}r)=1-e^{-a_{8}r}\sum _{i=0}^{2k}(a_{8}r)^{i}/i!
\end{equation*}
are the Tang-Toennies damping functions \cite{tang-toennies(1984)}. This potential function was originally proposed by Korona \textit{et al} \cite{korona-et-al(1997)}, and its extension was performed by Patkowski  \textit{et al} \cite{patkowski-et-al(2005)} to describe the \textit{ab initio} potential for argon dimer. Furthermore, Prudente \textit{et al} \cite{prudente-et-al(2009)} employed this potential function to adjusting \textit{ab initio} PECs for the diatomic molecules LiH and H$_{2}$. The numerical values of the $\{ a_{j}\} ,j=1,\dots ,8$ parameters and the dispersion coefficients, $C_{n},n=6,8,10$, obtained by Almeida \textit{et al} \cite{almeida-et-al(2011)} are given in Table \ref{tab parameters genetic algorithm}. In such paper they extended their previous methodology based on genetic algorithms \cite{marques_et_al(2008)} to fit the RbCs potential curve to spectroscopic data.

\begin{figure}[ht]
\begin{center}
	\includegraphics[scale=0.7,angle=0]{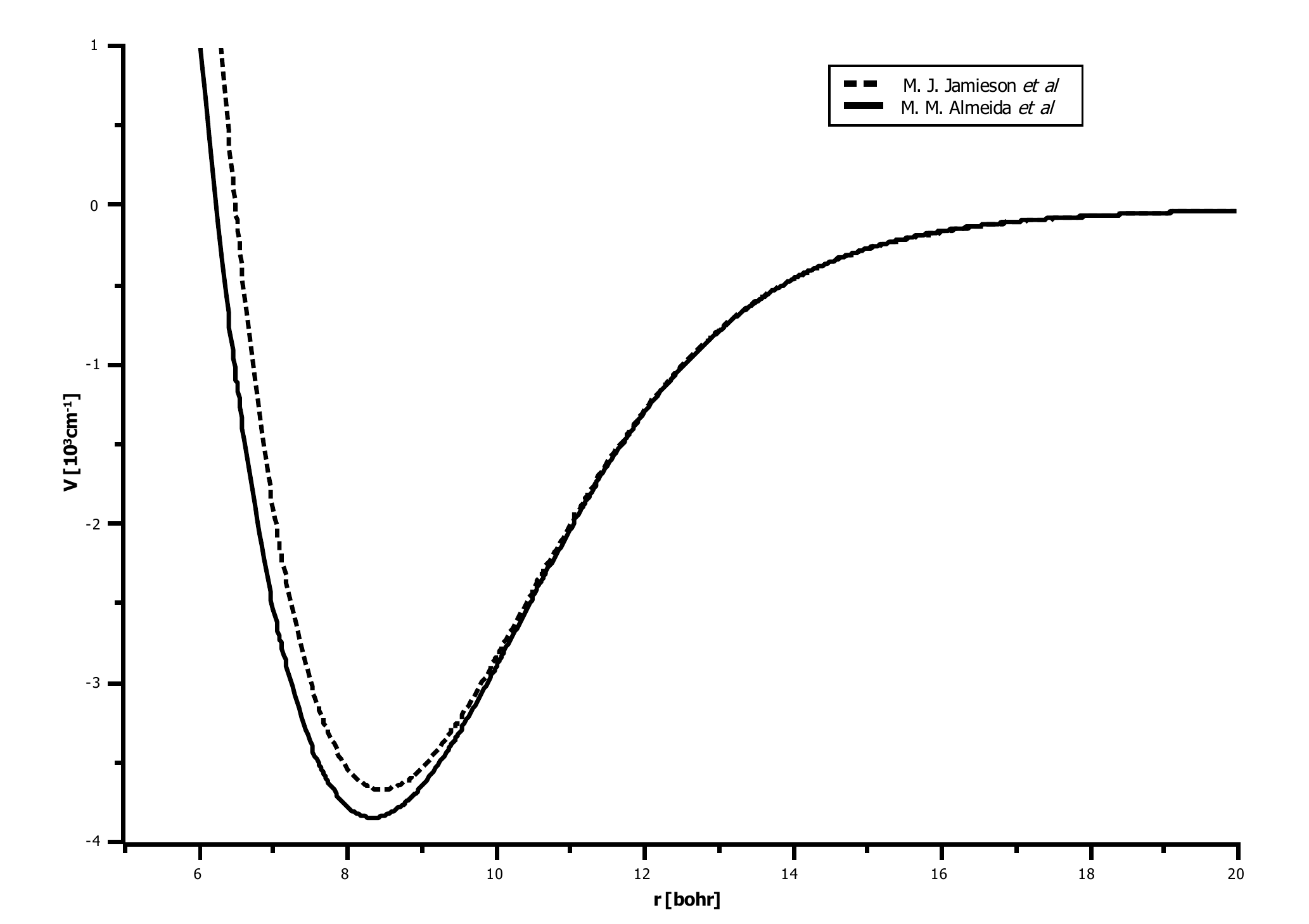}
\end{center}
	\caption{Potential energy curve for the $X^{1}\Sigma ^{+}$ state of RbCs molecule.}
	\label{fig pec rbcs}
\end{figure}

We also consider the potential proposed by Jamieson \textit{et al} \cite{jamieson-et-al(2003)} who used their \textit{ab initio} calculated short-range data matched at 17.9524 bohr to the analytical expression for long-range
\begin{equation}
V(r)=-\frac{C_{6}}{r^{6}}-\frac{C_{8}}{r^{8}}-\frac{C_{10}}{r^{10}}-Ar^{\beta }e^{-\gamma r}
\label{PEC Jamieson}
\end{equation}
with the parameters given in Table \ref{tab parameter jamieson}. In order to make a smooth connection between the two parts of the potential, it was joined the value of the long range potential, at 17.9524 bohr, the points \textit{ab initio} calculated utilizing an interpolation scheme of short-range potential by cubic spline \cite{press-et-al(1992)}. In the Figure \ref{fig pec rbcs} the PECs from equations \eref{PEC Almeida} and \eref{PEC Jamieson} are represented with the parameters given in Tables \ref{tab parameters genetic algorithm} and \ref{tab parameter jamieson}, respectively.

Jamieson \textit{et al} \cite{jamieson-et-al(2003)} used the Numerov's method to solve the radial Schr\"odinger equation for small asymptotic values of the wave number, $k$, determining the scattering length and effective range from the expansion \eref{effective range expansion}. Employing the potential proposed by them, shown in Figure \ref{fig pec rbcs}, they obtained the value of $a=40.24$ bohr, for the ${}^{85}$Rb$-{}^{133}$Cs  collision, and, $a=60.18$ bohr, for the ${}^{87}$Rb$-{}^{133}$Cs collision. In turn, Zanelatto \textit{et al} \cite{zanelatto-et-al(2005)} also used the Numerov's method to determine the scattering length. Employing the same potential, they obtained the value of $a=40.357$ bohr, for the ${}^{85}$Rb$-{}^{133}$Cs collision, and, $a=60.610$ bohr, for the ${}^{87}$Rb$-{}^{133}$Cs collision.

To achieving a good accuracy of our results, we have divided the FEM integration region in two intervals, one associated with the interaction region and another with the asymptotic region, and we have used an equidistant mesh in each intervals with $N_{e}=N^{in}_{e}+N^{out}_{e}$.
The scattering length as a function of the base parameters ($N_{e}$ and $n_{i}$) given in equation (\ref{fem expansion}) is represented by $a(N_{e},n_{i})$.
 Specifically, to ensure a convergence factor, $\Delta a=\vert a(N_{e},n_{i})-a(N_{e},n_{i}-1) \vert $, of at least five decimal places, we have used $N^{in}_{e}=60$ in the interval of $[0,40]$ bohr, $N^{out}_{e}=50$ in the interval of $[40,r_{max}]$ bohr and $n_{i}=30$, $\forall i $.
This convergence process is demonstrated, for ${}^{87}$Rb$-{}^{133}$Cs collision, in Table \ref{tab convergence process} where we have calculated $a$ and $\Delta a$ for different values of $N_{e}$ and $n_{i}$ keeping $N^{in}_{e}/N^{out}_{e}=6/5$ and $r_{max}=6000$ bohr, and employing the same potential as Jamieson \textit{et al} \cite{jamieson-et-al(2003)}.

The Table \ref{tab convergence process} also demonstrate that, for the energy $E=10^{-10}$ hartree, the scattering length calculated with the best base parameters has not yet converged in any decimal place to the same respective values calculated from the lower energies. It means that this energy value does not lead to a good approximation of equation (\ref{scattering length}). Thus, we consider the energy, $E$, of the order of $10^{-30}$ hartree ensuring a very small $k$ so that $a$ is calculated by the equation (\ref{scattering length}); this energy value is, for example, much lower than the one used by Zanelatto \textit{et al} \cite{zanelatto-et-al(2005)} who considered the energy of the order of $10^{-13}$ hartree.

Again, employing the same potential as Jamieson \textit{et al} \cite{jamieson-et-al(2003)}, we show in Table \ref{tab length x separation} the influence of maximum separation, $r_{max}$, in the convergence of the scattering length for the ${}^{85}$Rb$-{}^{133}$Cs and ${}^{87}$Rb$-{}^{133}$Cs collisions.
In the Table \ref{tab length x separation} we note that the present results converge to a value very close to those obtained by Jamieson \textit{et al} and Zanelatto \textit{et al} for a large maximum separation; the best agreement is reached in around $r_{max}=1000$ bohr, but continues its convergence as the maximum separation increases. Thereby, we evidence the efficiency of the present method to calculate the scattering length.

Now we consider the PEC, proposed by  Almeida \textit{et al} \cite{almeida-et-al(2011)}, with spectroscopic accuracy, obtained for $X^{1}\Sigma ^{+}$ state of RbCs, from a direct adjustment procedure of experimental data based on the genetic algorithm. It is given by equation \eref{PEC Almeida} and Table \ref{tab parameters genetic algorithm}. In Table \ref{tab scattering lenght data} we show the scattering length for ${}^{85}$Rb$-{}^{133}$Cs and ${}^{87}$Rb$-{}^{133}$Cs collisions. The present results were obtained using the FEM with $N_{e}=60$ in the interval $[0,40]$ bohr, $N_{e}=50$ in the interval $[40,6000]$ bohr and $n_{i}=30$. Also, in the table, results are shown for various PECs using several sets of short and long range data withdrawn of Refs. \cite{jamieson-et-al(2003)} and \cite{zanelatto-et-al(2005)}. Note that the scattering length is very sensitive to the PEC parameters. The maximum values found in the Table correspond to the set IV calculated by Jamieson \textit{et al} \cite{jamieson-et-al(2003)} using the iterated perturbation analysis (IPA) potential by Fellows \textit{et al} \cite{fellows-et-al(1983)} for the short-range interaction and the long-range data from the equation \eref{PEC Jamieson} smoothing to IPA potential. In turn, the minimum values of $a$ correspond to the set VIII calculated by Zanelatto \textit{et al} \cite{zanelatto-et-al(2005)} using the \textit{ab initio} short-range potential by Allouche \textit{et al} \cite{allouche-et-al(2000)}, the dispersion coefficients $C_{6}$ obtained in reference \cite{derevianko-et-al(2001)}, $C_{8}$ and $C_{10}$ obtained of reference \cite{marinescu-sadeghpour(1999)}, and the exchange parameters ($A$, $\beta $ and $\gamma $) obtained in reference \cite{allouche-et-al(2000)}. The same short-range data and dispersion coefficients of set VIII were used by Jamieson \textit{et al} \cite{jamieson-et-al(2003)} using the set VI, but with different exchange parameters. Zanelatto \textit{et al} \cite{zanelatto-et-al(2005)} also used the Fermi function to connect smoothly the terms of short and long range. It is notable that their results are the only ones that have a negative value, indicating a repulsive interaction between atoms.

\begin{figure}[ht]
\begin{center}
	\includegraphics[scale=0.35,angle=0]{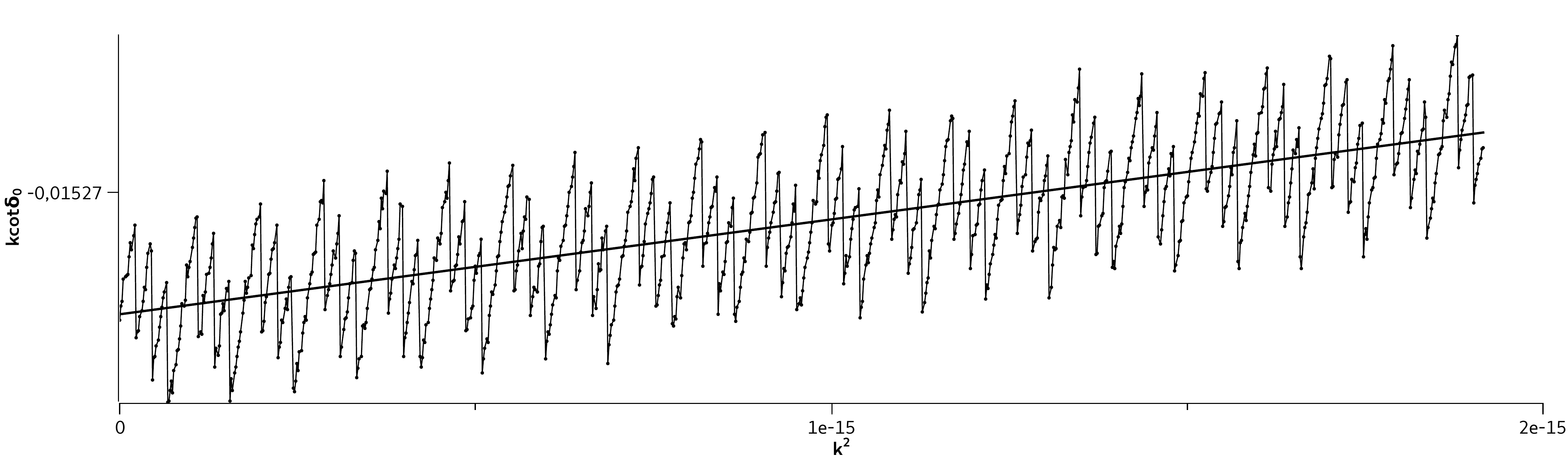}
\end{center}
	\caption{Graph of equation \eref{effective range expansion} describing of $k\cot \delta _{0}$ as function of $k^{2}$ for ${}^{87}$Rb$-{}^{133}$Cs collision. The energy $E=\hbar k^{2}/2\mu $ varies between $10^{-30}$ and $10^{-20}$ hartree. The straight line is the linear regression adjustment getting $k\cot \delta _{0}=-1.5269\cdot 10^{-2}+2.3667\cdot 10^{2}k^{2}$ (a.u.).}
	\label{fig effective range}
\end{figure}

We also determined the scattering length, for the PEC from Almeida \textit{et al}, using the effective range expansion \eref{effective range expansion}, describing $k\cot \delta _{0}$ as a function of $k^{2}$. This is shown in Figure \ref{fig effective range} for ${}^{87}$Rb$-{}^{133}$Cs collision. In order to maintain the results in concordance with the ones obtained by equation \eref{scattering length} we chosen a energy interval between $10^{-30}$ and $10^{-20}$ hartree. Then we have got a good estimative for the effective range obtaining $r_{c}=600.903$ bohr for ${}^{85}$Rb$-{}^{133}$Cs and $r_{c}=473.350$ bohr for ${}^{87}$Rb$-{}^{133}$Cs.

In particular, the results obtained with the PEC obtained with spectroscopic quality employing the FEM is closer of set VII, also calculated by  Jamieson \textit{et al} using their \textit{ab initio} short-range interaction potential and the theoretical values of the long-range parameters obtained in references \cite{marinescu-sadeghpour(1999),marinescu-dalgarno(1996)} but with $C_{6}$ replaced by very precise value of Derevianko \textit{et al} \cite{derevianko-et-al(2001)}. On the other hand, Almeida \textit{et al} \cite{almeida-et-al(2011)} determine the coefficients of multipolar electrostatic expansion of the interaction between the two atoms of the diatomic molecules comparing them with other values reported in the literature. They demonstrated that their results are those with the best agreement considering an experimental estimation of $\chi _{4}=C_{6}C_{10}/C_{8}^{2}$ close to $4/3$, as suggested by Le Roy \cite{leroy(1974)} based on the observation of the coefficients for electronic states of $\Sigma $ symmetry. Also the analysis of Thakkar \cite{thakkar(1988)} and Mulder \textit{et al} \cite{mulder-et-al(1980)} suggests a value of $\chi _{4}$ larger than $1.2$. This can be seen in Table \ref{tab multipolar coefficients} in which are presented the coefficients of multipolar electrostatic expansion taken from Table \ref{tab parameters genetic algorithm} and from several sets of Table \ref{tab scattering lenght data}. Thus, the $C_{6}$, $C_{8}$ and $C_{10}$ values of Almeida \textit{et al} indicate good estimate for the dispersion coefficients. This lead us to consider that the results obtained in the present study using the Almeida \textit{et al} PEC represent good estimates for scattering length and effective range of RbCs collision in ground state. 

\section{Conclusion}

In this paper we utilized a numerical procedure based on variational $R$-matrix and finite element methods to solve the radial Schr\"{o}dinger equation and perform the calculation of scattering length for the cold collision between the rubidium and cesium atoms. Also we test a potential energy curve recently fitted to spectroscopic data by Almeida \textit{et al} \cite{almeida-et-al(2011)} using a methodology based on genetic algorithm. We notice that our results agree with the previously published in literature. Whereas both variational principle and local basis functions are quite accurate methods for numerical solution of physical problems, we believe that the values displayed here, with the novel potential curve, can be a good estimation for scattering length and effective range of the ground electronic state of RbCs.

We pointed out that we utilize an efficient algorithm for obtaining the $R$-matrix based on a matrix inversion technique which has been successfully applied in other studies \cite{prudente-soares(1999),guimaraes-prudente(2011)}. This algorithm works with small block matrices and aims to achieve just the last block of the inverse matrix. As the generated matrices by our methodology is very sparse it is needed to keep into the computer's memory just few non-zero blocks. The advantage is that it reduces significantly both memory and computational effort required to invert the matrix, which is generally quite large in variational approaches.

\ack

This work was supported by the Brazilian agencies CNPq, CAPES and FAPESB.

\section*{References}

\bibliography{RbCs-Ref}

\begin{thebibliography}{10}

\bibitem{weiner(2003)}
J.~Weiner.
\newblock {\em Cold and Ultracold Collisions in Quantum Microscopic and
  Mesoscopic Systems}.
\newblock Cambridge University Press, Cambridge, 2003.

\bibitem{shimasaki-et-al(2015)}
T.~Shimasaki, M.~Bellos, C.~D. Bruzewicz, Z.~Lasner, and D.~DeMille.
\newblock Production of rovibrinic-ground-state rbcs molecules via
  two-photon-cascade decay.
\newblock {\em Phys. Rev. A}, 91:021401(R), 2015.

\bibitem{molony-et-al(2014)}
P.~K. Molony, P.~D. Gregory, Z.~Ji, B.~Lu, M.~P. K\"{o}ppinger, C.~R.~Le Sueur,
  C.~L. Blackley, J.~M. Hutson, and S.~L. Cornish.
\newblock Creation of ultracold $^{87}$rb$^{133}$cs molecules in the
  rovibrational ground state.
\newblock {\em Phys. Rev. Lett.}, 113:255301, 2014.

\bibitem{takekoshi-et-al(2014)}
T.~Takekpshi, L.~Reichs\" {o}llner, A.~Schindewolf, J.~M. Hutson, C.~R.~Le
  Sueur, O.~Dulieu, F.~Ferlaino, R.~Grimm, and H-C N\"{a}gerl.
\newblock Ultracold dense samples of dipolar rbcs molecules in the
  rovibrational and hyperfine ground state.
\newblock {\em Phys. Rev. Lett.}, 113:205301, 2014.

\bibitem{bruzewicz-et-al(2014)}
C.~D. Bruzewicz, M.~Gustavsson, T.~Shimasaki, and D.~DeMille.
\newblock Continuous formation of vibronic ground state rbcs molecules via
  photoassociation.
\newblock {\em New J. Phys.}, 16:023018, 2014.

\bibitem{zuters-et-al(2013)}
V.~Zuters, O.~Docenko, M.~Tamanis, R.~Ferber, V.~V. Meshkov, E.~A. Pazyuk, and
  A.~V. Stolyarov.
\newblock Spectroscopic studies of the $(4)^{1}\sigma ^{+}$ state of rbcs and
  modmodel of the optical cycle for ultracold $x^{1}\sigma ^{+}$ ($\nu =0$,
  $j=0$) molecule production.
\newblock {\em Phys. Rev. A}, 87:022504, 2013.

\bibitem{takekoshi-et-al(2012)}
T.~Takekoshi, M.~Debatin, R.~Rameshan, F.~Ferlaino, R.~Grimm, H-C N\"{a}gerl,
  C.~R.~Le Sueur, J.~M. Hutson, P.~S. Julienne, S.~Kotochigova, and E.~Tiemann.
\newblock Towards the production of ultracold ground-state rbcs molecules:
  Feshbach resonance, weakly bound states, and the coupled-channel model.
\newblock {\em Phys. Rev. A}, 85:032506, 2012.

\bibitem{yang-et-al(2012)}
Y.~Yang, X.~Liu, Y.~Zhao, L.~Xiao, and S.~Jia.
\newblock Rovibrational dynamics of rbcs on its lowest $^{1,3}\sigma ^{+}$
  potential curves calculated by couple cluster method with all-electron basis
  set.
\newblock {\em J. Phys. Chem.}, 116:11101, 2012.

\bibitem{docenko-et-al(2011)}
O.~Docenko, M.~Tamanis, R.~Ferber, H~Kn\"{o}ckel, and E.~Tiemann.
\newblock Singlet and triplet potentials of ground-state atom pair rb$+$cs
  studied by fourier-transform spectroscopy.
\newblock {\em Phys. Rev. A}, 2011:052519, 2011.

\bibitem{adhikari(1998)}
S.~K. Adhikari.
\newblock {\em Variational Principles and the Numerical Solution of Scattering
  Problems}.
\newblock John Willey \& Sons, New York, 1998.

\bibitem{wigner-eisenburg(1947)}
E.~P. Wigner and L.~Eisenbud.
\newblock Higher angular momenta and long range interaction in resonance
  reactions.
\newblock {\em Phys. Rev.}, 72:29, 1947.

\bibitem{burke(2011)}
P.~B. Burke.
\newblock {\em R-Matrix Theory of Atomic Collisions: Application to Atomic,
  Molecular and Optical Processes}.
\newblock Springer-Verlag, Berlin, 2011.

\bibitem{ram-moham(2002)}
L.~R. Ram-Moham.
\newblock {\em Finite Element and Boundary Element Applications in Quantum
  Mechanics}.
\newblock Oxford University Press, New York, 2002.

\bibitem{pask-et-al(2001)}
J.~E. Pask, B.~M. Klein, P.~A. Sterne, and C.~Y. Fong.
\newblock Finite-element methods in eletronic-structure theory.
\newblock {\em Comput. Phys. Commun.}, 135:1, 2001.

\bibitem{soares-prudente(1994)}
J.~J. {Soares Neto} and F.~V. Prudente.
\newblock A novel finite element method implementation for calculating bound
  states of triatomic systems: Application to the water molecule.
\newblock {\em Theor. Chim. Acta}, 89:415, 1994.

\bibitem{prudente-soares2(1999)}
F.~V. Prudente and J.~J. {Soares Neto}.
\newblock Optimized mesh for the finite-element method using a
  quantum-mechanical procedure.
\newblock {\em Chem. Phys. Lett.}, 302:43, 1999.

\bibitem{linderberg(1989)}
J.~Linderberg, S.~B. Padkj{\ae}r, Y.~{\"O}hrn, and B.~Vessal.
\newblock Numerical implementation of reactive scattering theory.
\newblock {\em J. Chem. Phys.}, 90:6254, 1989.

\bibitem{jaquet-schnupf(1992)}
R.~Jaquet and U.~Schnupf.
\newblock The $s$-matrix version of the hulth{\'e}n-kohn variational principle
  for quantum scattering: Comparison between conventional and finite element
  basis sets.
\newblock {\em Chem. Phys.}, 165:287, 1992.

\bibitem{bethe(1949)}
H.~A. Bethe.
\newblock Theory of the effective range in nuclear scattering.
\newblock {\em Phys. Rev.}, 76:38, 1949.

\bibitem{blatt(1948)}
J.~M. Blatt.
\newblock On the neutron-proton force.
\newblock {\em Phys. Rev.}, 74:92, 1948.

\bibitem{blatt-jackson(1949)}
J.~M. Blatt and J.~D. Jackson.
\newblock On the interpretation of neutron-proton scattering data by schwinger
  variational method.
\newblock {\em Phys. Rev.}, 76:18, 1949.

\bibitem{guimaraes-prudente(2005)}
M.~N. Guimar{\~a}es and F.~V. Prudente.
\newblock A study of the confined hydrogen atom using the finite element
  method.
\newblock {\em J. Phys. B: At. Mol. Opt. Phys.}, 38:2811, 2005.

\bibitem{prudente-soares(1999)}
F.~V. Prudente and J.~J. {Soares Neto}.
\newblock Quantum scattering using a novel implementation based on the
  variational $r$ matrix formalism and the finite element method: A comparative
  study.
\newblock {\em Chem. Phys. Lett.}, 309:471, 1999.

\bibitem{guimaraes-prudente(2011)}
M.~N. Guimar{\~a}es and F.~V. Prudente.
\newblock A variational adiabatic hyperspherical finte element $r$ matrix
  methodology: General formalism and application to h$+$h$_{2}$ reaction.
\newblock {\em Eur. Phys. J. D.}, 64:287, 2011.

\bibitem{almeida-et-al(2011)}
M.~M. Almeida, F.~V. Prudente, C.~E. Fellows, J.~M.~C. Marques, and F.~B.
  Pereira.
\newblock Direct fit of spectroscopic data of diatomic molecules by using
  genetic algorithms: Ii. the ground state of rbcs.
\newblock {\em J. Phys. B: At. Mol. Opt. Phys.}, 44:225102, 2011.

\bibitem{tang-toennies(1984)}
K.~T. Tang and J.~P. Toennies.
\newblock An improved simple model for the van der waals potential based on
  universal damping functions for dispersion coeffficients.
\newblock {\em J. Chem. Phys.}, 80:3726, 1984.

\bibitem{korona-et-al(1997)}
T.~Korona, H.~L. Williams, R.~Bukowski, B.~Jeziorski, and K.~Szalewicz.
\newblock Helium dimer potential from symmetry-adapted pertubation theory
  calculations using large gaussian geminal and orbital basis sets.
\newblock {\em J. Chem. Phys.}, 106:5109, 1997.

\bibitem{patkowski-et-al(2005)}
K.~Patkowski, G.~Murdachaew, C.-M. Fou, and K.~Szalewicz.
\newblock Accurate \textit{ab initio} potential for argon dimer including
  highly repulsive region.
\newblock {\em Mol. Phys.}, 103:2031, 2005.

\bibitem{prudente-et-al(2009)}
F.~V. Prudente, J.~M.~C. Marques, and A.~M. Maniero.
\newblock Time-dependent wave packet calculation of lih$+$h reactive scattering
  on new potential energy surface.
\newblock {\em Chem. Phys. Lett.}, 474:18, 2009.

\bibitem{marques_et_al(2008)}
J.~M.~C. Marques, F.~V. Prudente, M.~M. Almeida, and C.~E. Fellows.
\newblock A new genetic algorithm to be used in the direct fit of potential
  energy curves to ab initio and spectroscopic data.
\newblock {\em J. Phys. B: At. Mol. Opt. Phys.}, 41:085103, 2008.

\bibitem{jamieson-et-al(2003)}
M.~J. Jamieson, H.~{Sarbazi-Azad}, H.~Ouerdane, G.-H. Jeung, Y.~S. Lee, and
  W.~C. Lee.
\newblock Elastic scattering of cold caesium and rubidium atoms.
\newblock {\em J. Phys. B: At. Mol. Opt. Phys.}, 36:1085, 2003.

\bibitem{press-et-al(1992)}
W.~H Press, S.~A Teukolsky, W.~T. Vetterling, and B.~P. Flannery.
\newblock {\em Numerical Recipes in Fortran 77: The Art of Scientific
  Computing}.
\newblock Cambridge University Press, New York, 2 edition, 1992.

\bibitem{zanelatto-et-al(2005)}
A.~L.~M. Zanelatto, E.~M.~S. Ribeiro, and R.~d.~J.~Napolitano.
\newblock Scattering lengths for li$-$cs, na$-$cs, k$-$cs and rb$-$cs ultracold
  collisions.
\newblock {\em J. Chem. Phys.}, 123:014311, 2005.

\bibitem{fellows-et-al(1983)}
C.~E. Fellows, R.~F. Gutterres, A.~P.~C. Campos, J.~Verg{\`e}s, and C.~Amiot.
\newblock The rbcs $x^{1}\sigma ^{+}$ ground state: New spectroscopic study.
\newblock {\em Rep. Prog. Phys.}, 46:97, 1983.

\bibitem{allouche-et-al(2000)}
A.~R. Allouche, M.~Korek, K.~Fakherdding, A.~Chaalang, M.~Dagher, F.~Taher, and
  M.~Aubert-Fr{\'e}con.
\newblock Theoretical electronic structure of rbcs revisited.
\newblock {\em J. Phys. B: At. Mol. Opt. Phys.}, 33:2307, 2000.

\bibitem{derevianko-et-al(2001)}
A.~Derevianko, J.~F. Babb, and A.~Dalgarno.
\newblock High-precision calculations of van der waals coefficients for
  heteronuclear alkali-metal dimers.
\newblock {\em Phys. Rev. A}, 63:052704, 2001.

\bibitem{marinescu-sadeghpour(1999)}
M.~Marinescu and Sadeghpour.~H. R.
\newblock Long-range potentials for two-species alkali-metal atoms.
\newblock {\em Phys. Rev. A}, 59:390, 1999.

\bibitem{marinescu-dalgarno(1996)}
M.~Marinescu and A.~Dalgarno.
\newblock Analytical interaction potentials of the long range alkali-metal
  dimers.
\newblock {\em Z. Phys. D: At., Mol. Clusters}, 36:239, 1996.

\bibitem{leroy(1974)}
R.~J. {Le Roy}.
\newblock Long-range potential coefficients from rkr turning points: $c_{6}$
  and $c_{8}$ for $b$(${}^{3}\pi _{\mbox{ou}}{}^{+}$)-state cl$_{2}$, br$_{2}$,
  and i$_{2}$.
\newblock {\em Can. J. Phys.}, 52:246, 1974.

\bibitem{thakkar(1988)}
A.~J. Thakkar.
\newblock Higher dispersion coefficients: Accurate values for hydrogen atoms
  and simple estimates for other systems.
\newblock {\em J. Chem. Phys.}, 89:2092, 1988.

\bibitem{mulder-et-al(1980)}
F.~Mulder, G.~F. Thomas, and W.~J. Meath.
\newblock A critical study of some methods for evaluating the $c_{6}$, $c_{8}$
  and $c_{10}$ isotropic dispersion energy coefficients using the first row
  hydrides, co, co$_{2}$ and n$_{2}$o as models.
\newblock {\em Mol. Phys.}, 41:249, 1980.

\end{thebibliography}
\bibliographystyle{unsrt}

\newpage

\begin{table}[tb]
\caption{Parameters obtained by the genetic algorithm procedure of direct adjustment. The parameters $a_{i}$ are in atomic units.}
\centering
\begin{tabular}{ll}
\hline\hline
$a_{1}$ & $-$1.9504268 \\ 
$a_{2}$ & 0.395953461${}^{a}$ \\ 
$a_{3}$ & 8.2933763 \\ 
$a_{4}$ & $-$0.02599482 \\
$a_{5}$ & $-$0.00030692 \\
$a_{6}$ & 0.11351898 \\ 
$a_{7}$ & 0.03321360 \\ 
$a_{8}$ & 0.87509116 \\
$C_{6}$ \ ($\times 10^{6}\mbox{cm}^{-1}\mbox{\AA }^{6}$ ) & 29.783746 \\
$C_{8}$ \ ($\times 10^{8}\mbox{cm}^{-1}\mbox{\AA }^{8}$ ) & 11.085596 \\
$C_{10}$ ($\times 10^{10}\mbox{cm}^{-1}\mbox{\AA }^{10}$ ) & 4.8508464 \\
 \hline\hline
\multicolumn{2}{l}{${}^{a}$ {\small It is misspelled in reference \cite{almeida-et-al(2011)}.}}
\label{tab parameters genetic algorithm}
\end{tabular}
\end{table}

\begin{table}[tb]
\caption{Parameters (in atomic units) used by Jamieson \textit{et al} \cite{jamieson-et-al(2003)} for obtaing the analytical form of long-range potential.}
\centering
\begin{tabular}{ll}
\hline\hline
$C_{6}$ \ ($\times 10^{3}$) & 5.663 \\ 
$C_{8}$ \ ($\times 10^{5}$) & 7.3052 \\ 
$C_{10}$ ($\times 10^{7}$) & 10.831 \\ 
$A$ \ \ ($\times 10^{-3}$) & 1.5069 \\
$\beta $ & 5.5060 \\
$\gamma $ & 1.0797 \\ 
\hline\hline
\label{tab parameter jamieson}
\end{tabular}
\end{table}


\begin{table}
\centering
\caption{Scattering length, $a$, and, in parentheses, the convergence factor, $\Delta a(N_e,n_{i}) = |a(N_e,n_{i})-a(N_e,n_{i}-1)|$, results as function of basis definition ($N_e$ and $n_{i}$) being $N_e=N^{in}_{e}+N^{out}_{e}$ with $N^{in}_{e}/N^{out}_{e}=6/5$, and for $r_{max}=6000$ bohr. Calculated with the potential from Jamieson's \textit{et al} \cite{jamieson-et-al(2003)} (Figure \ref{fig pec rbcs}) for the ${}^{87}$Rb$-{}^{133}$Cs collision in the $X^{1}\Sigma ^{+}$ state. All the magnitudes in atomic units: $m({}^{133}\mbox{Cs})=132.905447$ a.u. and $m({}^{87}\mbox{Rb})=86.9091835$ a.u..}
\centering
\begin{tabular}{cclllll}
\hline\hline 
Energy & $N_{e}$ & \multicolumn{5}{c}{$a$ and $\Delta a$} \\ 
       &         & $n_i=15$ & $n_i=20$ & $n_i=25$ & $n_i=30$ & $n_i=35$ \\ \hline 
         &  88   & 14.738319 & 123.65673 & 72.010790  & 70.561921 &  70.561575 \\
         &       &                   & ($\approx 10^{2}$)  & ($\approx 10^{1}$) & ($\approx 10^{0}$) & ($\approx 10^{-4}$) \\ 
$10^{-10}$&   110     & -148.75215 & 74.153436 &  70.562123 & 70.561575  &  70.561575  \\
          &           &                     & ($\approx 10^{2}$) & ($\approx 10^{0}$) & ($\approx 10^{-4}$) & ($< 10^{-6}$) \\ 
         &  154   & 74.659203 & 70.562010 &  70.561576 &  70.561575 &  70.561575 \\
         &        &                    & ($\approx 10^{0}$) & ($\approx 10^{-4}$) & ($< 10^{-6}$) & ($< 10^{-6}$) \\ 
\hline
         &  88   & -6.4915552 & 6381.2704 & 61.630639 & 60.118525 & 60.118163 \\
         &       &                 & ($\approx 10^{3}$)  & ($\approx 10^{3}$) & ($\approx 10^{0}$) & ($\approx 10^{-4}$) \\ 
$10^{-15}$&   110     & 241.81065 & 63.849319 & 60.118736 & 60.118163 & 60.118163  \\
          &           &                     & ($\approx 10^{2}$) & ($\approx 10^{0}$) & ($\approx 10^{-4}$) & ($< 10^{-6}$) \\ 
         &  154   & 64.370015 & 60.118618 & 60.118163 & 60.118163 & 60.118163  \\
         &        &                    & ($\approx 10^{0}$) & ($\approx 10^{-4}$) & ($< 10^{-6}$) & ($< 10^{-6}$) \\ 
\hline
         &  88   & -6.4919374 & 6380.3086 & 61.630452  & 60.118336 &  60.117974 \\
         &       &                     & ($\approx 10^{3}$) & ($\approx 10^{3}$) & ($\approx 10^{0}$) & ($\approx 10^{-4}$) \\           
$10^{-20}$& 110 & 241.80990 & 63.849135 & 60.117974 & 60.117974  & 60.117974  \\
          &     &                    & ($\approx 10^{2}$) & ($\approx 10^{0}$) & ($< 10^{-6}$) & ($< 10^{-6}$) \\ 
         &  154   & 64.369832 & 60.118429 &  60.117975 &  60.117974 &  60.117974 \\
         &        &                    & ($\approx 10^{0}$) & ($\approx 10^{-4}$) & ($< 10^{-6}$) & ($< 10^{-6}$) \\ 
\hline
         &  88   & -6.4919374 & 6380.3085 &  61.630452 & 60.118336 &  60.117974 \\ 
         &      &                     & ($\approx 10^{3}$) & ($\approx 10^{3}$) & ($\approx 10^{0}$) & ($\approx 10^{-4}$) \\ 
$10^{-30}$& 110 & 241.80989 & 63.849135 & 60.118548 & 60.117974  & 60.117974 \\
          &     &                    & ($\approx 10^{2}$) & ($\approx 10^{0}$) & ($\approx 10^{-4}$) & ($< 10^{-6}$) \\ 
         &  154   & 64.369832 & 60.118429 & 60.117975  & 60.117974 &  60.117974 \\
         &        &                    & ($\approx 10^{0}$) & ($\approx 10^{-4}$) & ($< 10^{-6}$) & ($< 10^{-6}$) \\ 
\hline\hline 
\label{tab convergence process}
\end{tabular}
\end{table}


\begin{table}[tb]
\caption{Influence of the maximum separation, $r_{max}$, in convergence of the scattering length, $a$, both in bohr, for the Rb$-$Cs collision in the $X^{1}\Sigma ^{+}$ state, calculated with the potential from Jamieson's \textit{et al} \cite{jamieson-et-al(2003)} (Figure \ref{fig pec rbcs}). Mass, in atomic units: $m({}^{133}\mbox{Cs})=132.905447$ a.u.; $m({}^{85}\mbox{Rb})=84.9117893$ a.u.; $m({}^{87}\mbox{Rb})=86.9091835$ a.u..}
\centering
\begin{tabular}{lcc}
\hline\hline
& ${}^{85}$Rb$-{}^{133}$Cs & ${}^{87}$Rb$-{}^{133}$Cs \\ 
$r_{max}$ & $a$ & $a$ \\ 
\hline
200 & 67.6489 & 83.8008  \\ 
300 & 50.2289 &  69.5049 \\ 
400 & 44.5831 & 64.5238  \\ 
500 & 42.3731 & 62.5026  \\ 
600 & 41.3493 & 61.5461  \\ 
700 & 40.8135 & 61.0385  \\
800 & 40.5067 & 60.7450 \\
900 & 40.3188 & 60.5639  \\
1000 & 40.1972 & 60.4461 \\
1200 & 40.0579 & 60.3102 \\
1400 & 39.9862 & 60.2398 \\
1600 & 39.9456 & 60.1997 \\
1800 & 39.9208 & 60.1753 \\
2000 & 39.9050 & 60.1595 \\
6000 & 39.8634 & 60.1180 \\
\hline\hline
\label{tab length x separation}
\end{tabular}
\end{table}

\begin{table}[tb]
\caption{Scattering length , $a$, for various potential energy curves of $X^{1}\Sigma ^{+}$ state using several data sets. Mass, in atomic units: $m({}^{133}\mbox{Cs})=132.905447$; $m({}^{85}\mbox{Rb})=84.9117893$; $m({}^{87}\mbox{Rb})=86.9091835$.}
\centering
\begin{tabular}{lll}
\hline\hline
                      &${}^{85}$Rb$-{}^{133}$Cs & ${}^{87}$Rb$-{}^{133}$Cs \\ 
\hline                      
Present               & 59.0770                 & 65.4915 \\
Set I${}^{a}$    & 321.3                   & 417.1 \\
Set II${}^{a}$   & 115.9                   & 126.6 \\
Set III${}^{a}$  & 103.0                   & 112.4 \\
Set IV${}^{a}$   & 380.9                   & 564.2 \\
Set V${}^{a}$    & 27.79                   & 38.87 \\
Set VI${}^{a}$   & 0.0902                  & 12.43 \\
Set VII${}^{a}$  & 40.24                   & 60.18 \\
Set VIII${}^{b}$ & $-$40.618               & $-$11.125  \\
\hline\hline

\multicolumn{3}{l}{${}^{a}$ Jamieson \textit{et al} \cite{jamieson-et-al(2003)}} \\
\multicolumn{3}{l}{${}^{b}$ Zanelatto \textit{et al} \cite{zanelatto-et-al(2005)}} \\
\label{tab scattering lenght data}
\end{tabular}
\end{table}

\begin{table}[tb]
\caption{Coefficients of multipolar electrostatic expansion (in atomic units) taken from Table \ref{tab parameters genetic algorithm} and from several sets of Table \ref{tab scattering lenght data}.}
\centering
\begin{tabular}{lllll}
\hline\hline
                       & $C_{6}$  \ ($\times 10^{3}$) & $C_{8}$ \ ($\times 10^{5}$)   & $C_{10}$ \ ($\times 10^{7}$) & $\chi _{4}=C_{6}C_{10}/C_{8}^{2}$\\ 
\hline                      
Almeida \textit{et al} \cite{almeida-et-al(2011)} & 6.1800 & 8.2142 & 12.836 & 1.18 \\
Set I & 5.2840 & 7.3052 & 10.831 & 1.07 \\
Set II & 5.4785 & 8.566 & 11 & 0.82 \\
Set III & 5.4798 & 8.566 & 11 & 0.82 \\
Set IV & 5.4318 & 8.581 & 11 & 0.81 \\
Set V & 5.663 & 8.566 & 11 & 0.85 \\
Set VI--VIII & 5.663 & 7.3052 & 10.831 & 1.15 \\
\hline\hline
\label{tab multipolar coefficients}
\end{tabular}
\end{table}

\newpage

\end{document}